\documentstyle[eqsecnum,aps,preprint,epsfig]{revtex}
\begin{document}
\tightenlines
\draft
\title{Asymmetric Nuclear Matter in the Relativistic Brueckner 
Hartree Fock approach}

\draft

\author{F. de Jong and H. Lenske}
\address{Institut f\"ur Theoretische Physik, Universit\"at Giessen,
35392 Giessen, Germany}
\date{\today}

\maketitle
\begin{abstract}
We present a calculation of asymmetric nuclear matter properties
in a relativistic Brueckner Hartree Fock framework. 
Following other calculations the components of the self-energies 
are extracted by projecting on Lorentz invariant amplitudes. 
It is shown that for asymmetric nuclear matter one needs 
a sixth invariant. 
We present a set of invariants which in the limit of symmetric
nuclear matter reduces to the conventional set. 
We argue that the existence of such a properly behaving 
set is also crucial for the application of the projection
method in symmetric nuclear matter.
Results for the equation of state and other observables are 
presented.
Special attention is payed to an analysis in terms of mean-field
effective coupling constants. 
Apart from the usual ones we also find significant strength
in the isovector scalar channel, which can be interpreted as 
an effective $\delta$-meson. 
\end{abstract}

\pacs{21.65.+f, 21.60.-n}

\section{Introduction}

An important step in understanding the binding properties of nuclear
matter was made with relativistic approaches to nuclear interactions.
They were capable to resolve to a large extent the longstanding problem
of the underestimation of the binding energy of nuclear matter found in
non-relativistic calculations. Conceptually, the relativistic theory
provides  a very natural link to modern boson exchange interactions  and
- at least in principle - relates in-medium interactions to free
space nucleon-nucleon (NN) scattering.

Applications have mostly been concentrating on the description of
symmetric nuclear matter 
\cite{Anastasio,Machleidt,Horowitz,terHaar,FdJ_spec}.
Although being different in theoretical and numerical details the
various approaches are in rather close agreement on the bulk properties
of nuclear matter. Investigations of asymmetric nuclear matter, however,
are rare, most of them of a very recent date \cite{terHaar_asym,Engvik,Huber,Lee}. 
One reason is certainly that up to very recent years the data base on asymmetric 
systems was rather limited because stable nuclei range around $Z/A\simeq 0.5$ 
with a rather small bandwidth of variations. Asymmetric matter was mainly of
theoretical interest, except for astrophysical studies, as the structure
of neutron stars \cite{Pethick}. 
The situation has changed completely 
with the new radioactive beam facilities producing now new isotopes in a
sufficiently large amount for detailed studies. The isospin degree of
freedom in the nuclear equation of state can be investigated for the
first time over a much larger range of proton-neutron asymmetries as
before. At present, the largest asymmetries produced experimentally are
obtained for light dripline nuclei ($Z/A=0.2 \cdots 0.3$ for the heavy
isotopes of He to oxygen nuclides) which is about a factor of 2 less of
the asymmetries found in stable nuclei. In the near future an increasing
amount of data for nuclides all over the mass table can be expected. 

It is obvious that the properties of asymmetric matter are of high
interest for nuclear structure at the proton and neutron driplines and
{\em vice versa}. The isospin dependence of the nuclear equation of
state is to a large extent unexplored. Extrapolations of the nuclear
mass formula into the regions of new isotopes persistently result in
strong deviations from measured binding energies. The same is found for
structure calculations with seemingly well established effective
interaction models as the non-relativistic Skyrme energy functional. 
Although the parameters of Skyrme interactions are derived empirically 
and the various model interactions reproduce the properties of stable nuclei
very well the predictions for nuclei far off stability are found to differ
drastically. 
Similar observation are also made for relativistic
mean-field theories. Thus, the question arises to what extent the
asymmetry, i.e. isospin dependent contributions to the Bethe-Weisz\"acker mass
formula are really understood.

It is apparent that systematic studies of asymmetric matter are required.
In this work we describe asymmetric nuclear matter in the relativistic
Brueckner-Hartree-Fock (RBHF) approach. Compared to symmetric matter
the theoretical and numerical efforts, respectively, are much larger because
protons and neutrons are occupying different Fermi spheres. 
This implies different effective masses for protons and neutrons and the 
two sorts of particles cover a different range of off-shell momenta. 
Hence, it can be expected that beside the isospin also the momentum dependence of
self-energies will be of importance.
As noted before, asymmetric matter requires
to solve a system of coupled equations for the set 
$T_{\tau_1,\tau_2,\tau_3,\tau_4}$ of in-medium proton-neutron T-matrices, 
where $\tau=\pm 1$ denote a proton or neutron, respectively. 
We perform our calculation in a plane-wave spinor basis in which we 
have neutrons and protons as distinguishable particles.
This implies that we also take into account partial-wave amplitudes
that are zero in symmetric nuclear and neutron matter, where the 
masses are equal and isospin is a good quantum number.
It is not clear whether other calculations take these amplitudes into
account.
Details of the theoretical formulation are discussed in Sect. 2.

A central point of the discussion is the derivation of a new asymmetry
dependent invariant. Traditionally, the positive energy-projected
in-medium on-shell T-matrix is expressed in terms of five Lorentz
invariant amplitudes (see Eq.(2.7)). This standard set of invariants is
known to be not unique because each can be replaced by its derivative
counterpart. A well known example is the ambiguity found by changing
from the pseudo-scalar to the pseudo-vector $\pi-NN$ coupling. A
proper treatment avoiding such ambiguities is to solve the in-medium
scattering problem for the full Dirac-space including also negative
energy states. Such an approach was applied by Huber et al. \cite{Huber}
using the so-called $\Lambda$-approximation. 
In our approach we continue to use the projected T-matrix.
To do so we need to introduce a sixth invariant since in asymmetric 
matter there are six independent helicity amplitudes. 
Clearly, the new invariant must be defined such that the description of
{\em symmetric} matter is conserved. A promising guideline is to consider
an amplitude which approaches zero for equal proton and neutron
effective masses. Such an amplitude will vanish in symmetric matter and
thus fulfills the constraint mentioned above. In order to account for
the differences in $T_{np,np}$ and $T_{np,pn}$ one actually has to
define separate invariant amplitudes for the direct and the exchange
channel.  In the direct channel a vector-scalar and in the exchange
channel a pseudovector-pseudoscalar combination is required.  However,
only the direct amplitude contributes to the proton/neutron
self-energies.

Results for the equation of state for nuclear matter at various degrees
of asymmetry are discussed in Sect. 3. Particular emphasis is put on the
momentum structure of self-energies. We define effective isoscalar and
isovector self-energies which directly characterize the mean-field of
asymmetric nuclear matter. The dependence of the equation of state on
the isovector density is tested and compared to the prediction of the
mass formula of being quadratic. The paper closes in Sect. 4 with a
summary and conclusions.

\section{Theoretical Framework}

The implementation of the RBHF model we use is identical to the one
presented in Ref. \cite{terHaar}, an account of a calculation for
asymmetric nuclear matter can be found in Ref. \cite{terHaar_asym}.
We will only briefly review the concepts of the relativistic Brueckner
model, but will elaborate further on an omission in the
calculation of Ref. \cite{terHaar_asym}.

The central quantity in the model is the self-energy,
in the RBHF model it has a structure in the spinor representation, 
expressed by scalar and vector components:
\begin{equation}
\Sigma(p) = \Sigma^s(p) - \gamma^0 \Sigma^0(p) + \bar{\gamma} \cdot \bar{p} \Sigma^v.
\end{equation}
Inserting this form of the self-energy into the Dirac equation, it is natural to 
define 'effective' quantities by 
\begin{equation}
m^* = m_N + \Sigma^s(p),\, p_0^* = p_0 + \Sigma^0(p),\, \bar{p}^* = (1 + \Sigma^v(p)) \bar{p}.
\end{equation}
This can be simplified further by dividing out the (small) $\Sigma^v$ 
term \cite{Horowitz,terHaar}:
$m^* = m_N + \Sigma^s(p)/(1 + \Sigma^v(p))$.
One thus finds a solution of the 'dressed' Dirac equation; the effective spinor:
\begin{equation}
u^*_r(\bar{p}) = {\left(\frac{E^*_p + m^*}{2m^*}\right)}^{\frac{1}{2}}
{
\left(
 \begin{array}{c}
 1 \\
 \displaystyle{\frac{\bar{\sigma}\cdot \bar{p}}{E^*_p + m^*}}
 \end{array}
\right)}
\chi_r,
\label{effspinor_ch3}
\end{equation}
with $E^* = \sqrt{\bar{p}^2 + {m^*}^2}$.
The self-energy is found to depend only weakly on the momentum at least inside 
the Fermi-sphere: in the practical
calculation it is taken constant and set to the value at the Fermi-surface. 
This greatly simplifies the calculations.

Obviously in asymmetric nuclear matter the Fermi-momenta of protons and
neutrons are different, leading to different Pauli-blocking operators and
corresponding neutron and the proton self-energies. 
This asymmetry is also reflected in the fact that one has
three different effective interactions, 
represented by three different $T$-matrices: $T_{pp,pp}$ (proton-proton scattering), 
$T_{pn,pn/np}$ (proton-neutron scattering) and $T_{nn,nn}$ (neutron-neutron scattering).
In the Brueckner scheme these are calculated by:
\begin{eqnarray}
T_{nn,nn} &=& V_{nn,nn} + \int V_{nn,nn} g_{nn} T_{nn,nn} \nonumber \\
T_{pp,pp} &=& V_{pp,pp} + \int V_{pp,pp} g_{pp} T_{pp,pp} \nonumber \\
T_{np,np} &=& V_{np,np} + \int V_{np,np} g_{np} T_{np,np} 
+ \int V_{np,pn} g_{pn} T_{pn,np}\nonumber \\
T_{pn,np} &=& V_{pn,np} + \int V_{pn,np} g_{np} T_{np,np} 
+ \int V_{pn,pn} g_{pn} T_{pn,np}
\end{eqnarray}
Note the coupled channels for $T_{np}$, we also stress that $V_{np,pn}$ is {\it not}
related by a Fierz-transform to $V_{np,pn}$: in $V_{np,pn}$ we have 
neutron-proton vertices on which only the isospin-1 mesons contribute. 
It immediately follows that the same holds for the $T$-matrices: 
$T_{np,np}$ is not related to $T_{np,pn}$.
The 2-body propagators $g_{ij}$ are the usual ones as they appear in the 
three-dimensionally reduced Thompson equation:
\begin{equation}
g_{ij} = \frac{m^*_i}{E^*_i} \frac{m^*_j}{E^*_j} 
\frac{\bar{Q}_{ij}(P,\sqrt{s^*},p)}{\sqrt{s^*} - E^*_i - E^*_j + i\epsilon},
\end{equation}
where $\bar{Q}_{ij}$ is the angle averaged Pauli blocker of the intermediate 
momentum $p$,
a function of the total momentum $P$ and invariant mass $\sqrt{s^*}$,
its parameters depend on the appropriate Fermi-momenta $p_{f,i/j}$.
After obtaining a solution for the $T$-matrices, the self-energy
is then calculated using a Brueckner-type expression:
\begin{equation}
\Sigma_N = \int Tr \left[ T^{A}_{nn} g_n \right] + 
\int Tr \left[ T^{A}_{np} g_p \right],
\end{equation}
where $T^{A}_{np}$ stands for $T_{np,np}(\theta = 0) - T_{np,pn}(\theta = \pi)$.
For the proton self-energy the expression is completely similar.
We will specify these expressions further later on.
For the moment we note that the integrations extend over the neutron and 
proton Fermi sea, respectively. 
This calculation is repeated until a self-consistent solution for the self-energy
is found.
With the self-consistent self-energy one then can calculate quantities like 
binding energy and single-particle energies.  

In our implementation of the relativistic Brueckner model we calculate the
self-energy by decomposing the various $T$-matrices in Lorentz invariant 
amplitudes \cite{Horowitz}.
This boils down to transforming the spin-representation to spinor-representation.
The contributions of the various amplitudes to the components of the self-energy
are then easily found. 
In the case of identical particles (to be more precise, particles with 
equal mass), five amplitudes are needed for the decomposition of the on-shell
$T$-matrix. 
We use a set of invariants including the pseudo-vector coupling. 
In the c.m. system we then have:
\begin{eqnarray}
\langle p' | T(P,\sqrt{s^*}) | p \rangle &=& 
\sum_i \Gamma_i(P,\sqrt{s^*},p',p) \Phi_i \nonumber \\
\Phi_i &=& \left[\bar{u}^*(\bar{p}') f_i u^*(\bar{p}) \right]
\left[\bar{u}^*(-\bar{p}') f_i u^*(-\bar{p}) \right] \nonumber \\
f_i &\in& \{ 1, \gamma^\mu, \sigma^{\mu \nu}, \gamma^5\gamma^\mu, 
\frac{\mbox{$\not \! q$} \gamma^5}{2 m^*} \}.
\end{eqnarray}  
With $q$ the transferred momentum: $q = p' - p$.
For the on-shell scattering matrix elements we have in the c.m. frame
$|p'| = |p|$ and $\sqrt{s^*} = E^*_i + E^*_j$. 
To calculate the self-energy one needs the amplitudes for the 
direct ($\theta = 0, \bar{p}' = \bar{p}$) and exchange 
($\theta = \pi, \bar{p}' = -\bar{p}$) seperately.
At these points the set of invariants is degenerate, this is solved by decomposing
at small angles and extrapolating to $\theta = 0, \pi$. 
In Ref. \cite{Horowitz} it is shown that this limit exists.

It has often been noted this particular set of invariants is not unique. 
One can always replace an invariant with its derivative counterpart,
we did this by using the pseudo-vector invariant instead of the 
pseudo-scalar invariant. 
In this specific case one finds the same coefficient $\Gamma_i$ since the 
$T$-matrix is on-shell, but the contribution to the self-energy
is different. 
However, the contribution to the total energy (found by sandwiching the 
self-energy between effective spinors) is the same for a given $T$-matrix.
In a self-consistent calculation there will be a difference due to the fact that
the self-consistent self-energy will be different, and thus the $T$-matrix is
different.
One guideline to limit the freedom in the choice of invariants is to
require that one uses the
same vertices in the invariant amplitudes as appear in the OBE interaction.
Since we have a pseudo-vector coupling for the pion, we thus also
have to use the pseudo-vector invariant in the projection.

To avoid this ambiguity one has to calculate the scattering in the 
complete Dirac space, as is done e.g. in Ref. \cite{Huber}.
For the unique decomposition one needs 
$\bar{u}^*(\bar{p}) \Sigma(\bar{p}) u^*(\bar{p})$,
$\bar{u}^*(\bar{p}) \Sigma(\bar{p}) v^*(\bar{p}) = $
$\bar{v}^*(\bar{p}) \Sigma(\bar{p}) u^*(\bar{p})$ and
$\bar{v}^*(\bar{p}) \Sigma(\bar{p}) v^*(\bar{p})$.
In our calculation we only have the amplitude between positive energy
spinors (the mean-field generated by the self-energy). 
Two different decompositions of the self-energy $\Sigma,\Sigma'$
need to give the same value for the mean-field: 
$\bar{u}^*(\bar{p}) \Sigma(\bar{p}) u^*(\bar{p}) = 
\bar{u}^*(\bar{p}) \Sigma'(\bar{p}) u^*(\bar{p})$.
One can easily check that up to ${\cal O}(p^4/{m^*}^4)$ this implies
for the components
$\Sigma_s' = \Sigma_s - \alpha$,  
$\Sigma_0' = \Sigma_0 - \alpha$ and  
$\Sigma_v' = \Sigma_v - \alpha/2m^*$.
The `ambiguity' coefficient $\alpha$ will of course depend on density and 
momentum.
Inserting this in the expression for the renormalized effective mass
we see that this is shifted by a quantity much smaller than $\alpha$. 
Typically, an error of 100 MeV leads to a shift of $\sim 40$ MeV in the 
renormalized effective mass. 
From the above it follows that any approximate method for the 
decomposition of the self-energy will still give fair results for 
quantities that involve only the mean-field, like the binding energy.

In the case of particles having different masses, the helicity matrix elements 
cease to be symmetric under exchange of particles. 
Since the number of invariants needed to decompose the $T$-matrix is equal to
the number of independent helicity matrix elements after exhausting all
the symmetries, the loss of this symmetry implies that one needs 
additional invariants to decompose the $T$-matrix.
Appearantly this was not considered in Ref. \cite{terHaar_asym}.
For the on-shell $np$ $T$-matrix, needed for the calculation of
the self-energy, six invariants are required.
Although it is relatively easy to find additional invariants that
span the six-dimensional space of the six independent helicity matrix 
elements, the problem is that we need to have a decomposition that reduces 
to the one used in the symmetric case.
With this we mean that the coefficient of the additional, sixth, amplitude 
should approach zero in the limit of equal masses. 
Obviously this is not a trivial requirement. 

Since $T_{np,np}$ and $T_{np,pn}$ are different
we can choose invariants for the exchange and direct
amplitudes independently.
We encountered a similar situation for the $N\Delta$ scattering matrix
\cite{FdJ_delta}.
For the present case we found the following invariants which satisfy 
the above criterium:    
\begin{eqnarray}
\Phi_6^{dir} &=& 
\left[\bar{u}^*_n(\bar{p}') \mbox{$\not \! k$} u^*_n(\bar{p}) \right]
\left[\bar{u}^*_p(-\bar{p}') u^*_p(-\bar{p}) \right] +
\left[\bar{u}^*_n(\bar{p}') u^*_n(\bar{p}) \right]
\left[\bar{u}^*_p(-\bar{p}') \mbox{$\not \! k$} u^*_p(-\bar{p}) \right], \nonumber\\
\Phi_6^{exch} &=& 
\left[\bar{u}^*_n(\bar{p}') \gamma^5 \mbox{$\not \! k$} u^*_p(\bar{p}) \right]
\left[\bar{u}^*_p(-\bar{p}') \gamma^5 u^*_n(-\bar{p}) \right] +
\left[\bar{u}^*_n(\bar{p}') \gamma^5 u^*_p(\bar{p}) \right]
\left[\bar{u}^*_p(-\bar{p}') \gamma^5 \mbox{$\not \! k$} u^*_n(-\bar{p}) \right],
\end{eqnarray}
where we used standard definitions from scattering theory $k = p_1 - p_4$ which 
results in $k_0 = E_1 - E_4, \bar{k} = \bar{p} + \bar{p}'$.
The numerical indices are the particle numbers (1 and 2 are incoming, 3 and 4 outgoing), 
the indices $n,p$ denote a neutron and a proton spinor respectively.
The amplitude in the exchange channel is antisymmetric under exchange of particles,
even when the masses are equal (the other invariants are symmetric under exchange of
particles).
This does not mean that it is not a valid invariant, but it results in a vanishing
coefficient for equal masses, which is exactly what we required.
We were unable to find an invariant with this property for the direct channel,
but the one given above does have a vanishing coefficient for equal masses. 
In both invariants the momentum $k$ appears. 
This makes it impossible to interpret them as some sort of meson channel since
a meson carries a momentum $q = p_1 - p_3$.
To us there is no straightforward interpretation of these invariants beyond 
that they are needed to project out the $T$-matrix in the $np$ channel.
This $T$-matrix spans up a six-dimensional space in spin-space, and we therefore 
need a six-dimensional basis in spinor space for the basis transformation from spin
to spinor representation.
We want to stress again that it is crucial for the description of 
{\em symmetric} nuclear matter that for asymmetric nuclear matter we have a set 
of six invariants that behaves properly in the limit of symmetric nuclear matter
(equal proton and neutron masses). 
If we could not find such a set, the method of extracting self-energies in
symmetric nuclear matter by means of invariant amplitudes would have
an additional ambiguity since the self-energies would not be a 
continuous function of the asymmetry parameter.

Inserting the decomposition in the expression for the
self-energies we find:
\begin{eqnarray}
\Sigma_n(p_{f,n}) &=& 
\int_0^{p_{f,n}} \frac{d^3q}{(2\pi)^3} \frac{1}{2E^*_{q,n}}
\sum_i 
\left\{
f^i Tr[ (\not\!q^* + m^*_n) f^i] \Gamma^i_{dir,nn} -
f^i (\not\!q^* + m^*_n) f^i \Gamma^i_{exch,nn}
\right\}
\nonumber\\
&+& 
\int_0^{p_{f,p}} \frac{d^3q}{(2\pi)^3} \frac{1}{2E^*_{q,p}}
\sum_i
\left\{
f^i Tr[ (\not\!q^* + m^*_p) f^i] \Gamma^i_{dir,np} -
f^i (\not\!q^* + m^*_p) f^i \Gamma^i_{exch,np}
\right\}.
\end{eqnarray}
Working out the Dirac algebra, we find the
components of the self-energy expressed in terms of the various
$\Gamma^\alpha(s^*,P)$. 
Taking the Fermi-momentum $p_f$ along the $z$-axis and transforming to the
c.m.-defining variables $s^* = (p_1 + p_2)^2$ and $P = \bar{p}_1 + \bar{p}_2$
we have for the scalar and vector components of the self-energy
at the Fermi-surface
\begin{eqnarray}
\lefteqn{\Sigma^s_n(p_{f,n}) =}\nonumber\\ 
& &\int_{0}^{4 p_{f,n}^2} dP^2
 \int_{s^*_{min, n}}^{s^*_{max, n}} d s^* 
 \frac{1}{32 \pi^2 p_{f,n}} \frac{m^*_n}{E^*_{p_{f,n}} + E^*_{q,n}}
\nonumber\\
& &\hspace{1.5mm} \times (4\Gamma^{s}_{dir,nn} 
 - \Gamma^{s}_{exch,nn} - 4\Gamma^{v}_{exch,nn} 
 - 12\Gamma^{t}_{exch,nn} + 4\Gamma^{av}_{exch,nn} 
 + \frac{{m^*_n}^2 - p_{f,n}^\mu q_\mu}{2 {m^*_n}^2} \Gamma^{pv}_{exch,nn}) + \nonumber\\
& &\int_{0}^{4 p_{f,p}^2} dP^2
 \int_{s^*_{min, p}}^{s^*_{max, p}} d s^* 
 \frac{1}{32 \pi^2 p_{f,n}} \frac{m^*_p}{E^*_{p_{f,n}} + E^*_{q,p}}
\nonumber\\
& &\hspace{1.5mm} \times (4\Gamma^{s}_{dir,np}
 + 4\frac{p_{f,n}^\mu q_\mu - {m^*_p}^2}{m^*_p} \Gamma^6_{dir,np}
 - \Gamma^{s}_{exch,np} - 4\Gamma^{v}_{exch,np} 
 - 12\Gamma^{t}_{exch,np} + 4\Gamma^{av}_{exch,np} \nonumber\\
& &\hspace{1.5mm} \hphantom{\times (}
 + \frac{2({m^*_p}^2 - p_{f,n}^\mu q_\mu) - {m^*_p}^2 + {m^*_n}^2}{(m^*_n + m^*_p)^2}
   \Gamma^{pv}_{exch,np}),
\label{sig_s}\\
\vspace{3mm} & & \nonumber\\
\lefteqn{\Sigma^0_n(p_{f,n}) =}\nonumber\\ 
& &\int_{0}^{4 p_{f,n}^2} dP^2
 \int_{s^*_{min, n}}^{s^*_{max, n}} d s^* 
 \frac{1}{32 \pi^2 p_{f,n}} \frac{E^*_{q,n}}{E^*_{p_{f,n}} + E^*_{q,n}}
\nonumber\\
& &\hspace{1.5mm} \times (-4\Gamma^{v}_{dir,nn}
 + \Gamma^{s}_{exch,nn} - 2\Gamma^{v}_{exch,nn}
 - 2\Gamma^{av}_{exch,nn} 
- \frac{E^*_{p_{f,n}}}{E^*_q} \frac{{m^*_n}^2 - p_{f,n}^\mu q_\mu}{2 {m^*_n}^2} 
\Gamma^{pv}_{exch,nn}) + \nonumber \\
& &\int_{0}^{4 p_{f,p}^2} dP^2
 \int_{s^*_{min, p}}^{s^*_{max, p}} d s^* 
 \frac{1}{32 \pi^2 p_{f,n}} \frac{E^*_{q,p}}{E^*_{p_{f,n}} + E^*_{q,p}}
\nonumber\\
& &\hspace{1.5mm} \times (-4\Gamma^{v}_{dir,np} 
 - 4 m^*_p \frac{E^*_{p_{f,n}} - E^*_{q,p}}{E^*_{q,p}} \Gamma^6_{dir,np}
 + \Gamma^{s}_{exch,np} - 2\Gamma^{v}_{exch,np} - 2\Gamma^{av}_{exch,np} 
\nonumber\\
& &\hspace{1.5mm} \hphantom{\times (}
 - \frac{2E^*_{p_{f,n}}({m^*_p}^2 - p_{f,n}^\mu q_\mu) - E^*_{q,p}({m^*_p}^2 - {m^*_n}^2)}
{E^*_{q,p}(m^*_n + m^*_p)^2}
\Gamma^{pv}_{exch,nn}) 
\label{sig_0} \\
\vspace{3mm} & & \nonumber\\
\lefteqn{\Sigma^v_n(p_{f,n}) =}\nonumber\\
& &\int_{0}^{4 p_{f,n}^2} dP^2
 \int_{s^*_{min, n}}^{s^*_{max, n}} d s^* 
 \frac{1}{32 \pi^2 p_{f,n}} \frac{m^*_n}{E^*_{p_{f,n}} + E^*_{q,n}}
\nonumber\\
& &\hspace{1.5mm} \times (-4\Gamma^{v}_{dir,nn}
 + \Gamma^{s}_{exch,nn} - 2\Gamma^{v}_{exch,nn}
 - 2\Gamma^{av}_{exch,nn} 
- \frac{p_{f,n}}{q_{\rm z}} \frac{{m^*_n}^2 - p_{f,n}^\mu q_\mu}{2 {m^*_n}^2} 
\Gamma^{pv}_{exch,nn}) + \nonumber \\
& &\int_{0}^{4 p_{f,p}^2} dP^2
 \int_{s^*_{min, p}}^{s^*_{max, p}} d s^* 
 \frac{1}{32 \pi^2 p_{f,n}} \frac{q_z}{E^*_{p_{f,n}} + E^*_{q,p}}
\nonumber\\
& &\hspace{1.5mm} \times (-4\Gamma^{v}_{dir,np} 
- 4 m^*_p \frac{p_{f,n} - q_z}{q_z} \Gamma^6_{dir,np}
+ \Gamma^{s}_{exch,np} - 2\Gamma^{v}_{exch,np} - 2\Gamma^{av}_{exch,np}
\nonumber\\
& &\hspace{1.5mm} \hphantom{\times (} 
- \frac{2{p_{f,n}}({m^*_p}^2 - p_{f,n}^\mu q_\mu) - q_z({m^*_p}^2 - {m^*_n}^2)}
{q_z(m^*_n + m^*_p)^2}
\Gamma^{pv}_{exch,np}).
\label{sig_v}
\end{eqnarray}%
$q$ is the momentum of the integrated particle in the 
nuclear matter rest-frame, found by
applying the inverse transform to the set ($s^*,P$)
Also $p^\mu_f$ is the 
on-shell four-momentum of the incoming particle, taken at the
Fermi momentum, so $p^0_{f,i} = E^*_i$.
The integration limits are given by 
$s^*_{min,i} = (E^*_{p_{f,i}} + E^*_{q_{min,i},i})^2 - P^2$ with
$q_{min,i} = |P - p_{f,i}|$ and $s^*_{max,i} = 4{E^*_{p_{f,i}}}^2 - P^2$.
The index $i$ stands for either a neutron or a proton; 
$p_{f,i}$ are the respective Fermi-momenta and 
$E^*_{p,i} = (p^2 + {m^*_i}^2)^{1/2}$.

The coefficient $\Gamma^6_{exch,np}$ does not appear in the expressions for 
the self-energy: for $\theta = \pi$, $\bar{k} = 0$ and since the 
extrapolated $\Gamma^6_{exch,np}$ is finite (we checked this by 
renormalizing the invariant by dividing by $|k|$) it does not contribute 
to the self-energy.

\section{Results}

\begin{figure}
\centerline{\epsfig{file=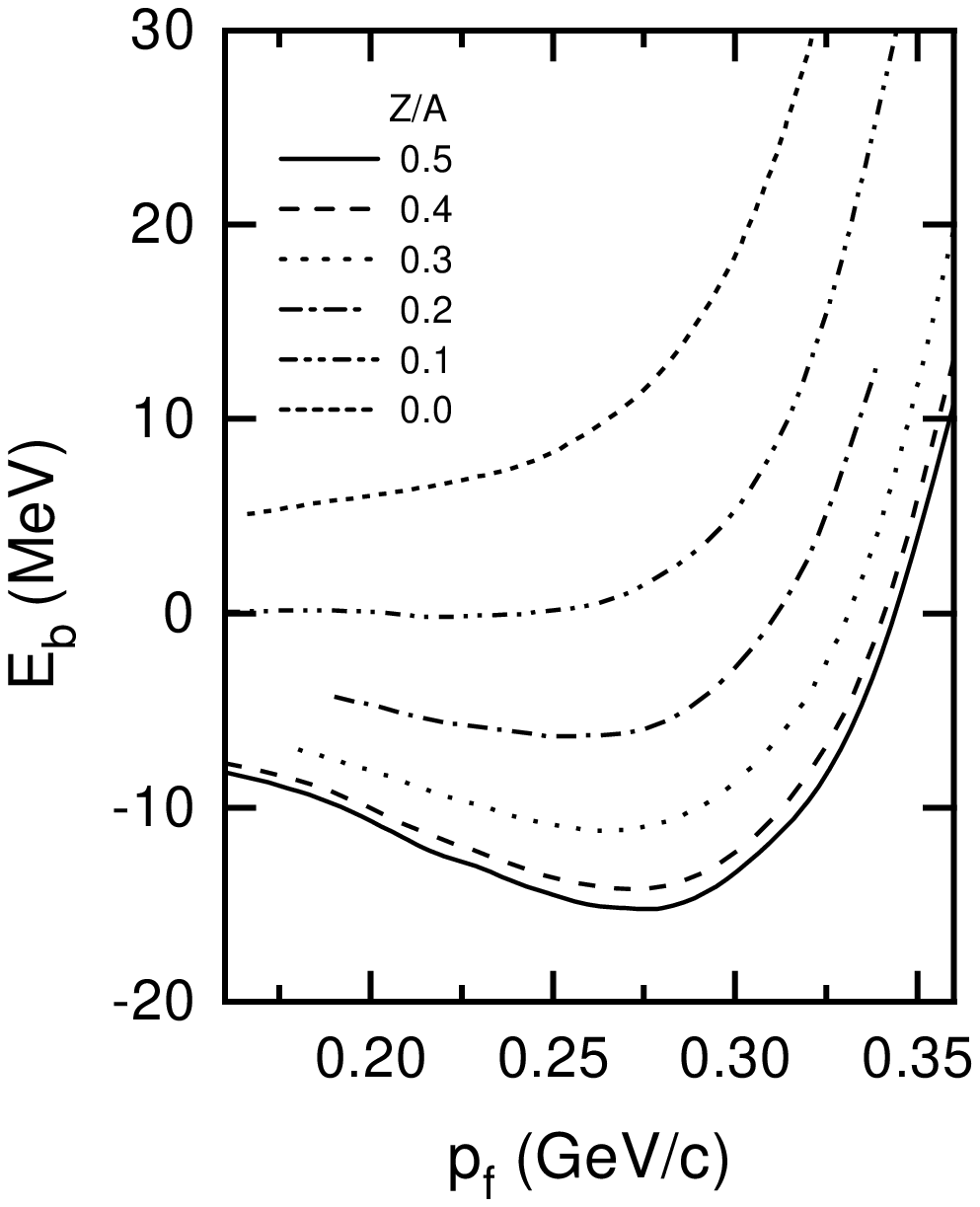,height=20cm}}
\caption{
The binding energy as a function of the density for proton ratios
$Z/A$ ranging from 0 (neutron matter) to 0.5 (nuclear matter).
}
\label{Figure_Eb}
\end{figure}

In Fig. \ref{Figure_Eb} we present the results for the equation of state for 
various values of the asymmetry parameter $Z/A$ using the Groningen 
potential \cite{terHaar}.
Following Ref. \cite{terHaar_asym} we used an 'averaged' Fermi momentum
defined by:
\begin{equation}
\rho = \frac{2}{3\pi^2} \bar{p}_f^3 = \frac{1}{3\pi^2}{p_{f,p}^3} 
+ \frac{1}{3\pi^2}{p_{f,n}^3}.
\label{fermi_mom}
\end{equation}
In the calculation it turns out that the contribution of the sixth invariant
is negligible, although it of course indirectly affects the results by influencing
the values of the other coefficients.
Overall the results look pretty much like the ones presented in Refs. 
\cite{terHaar_asym,Engvik,Huber,Lee}.
In a way this is not surprising because the results for $Z/A = 0.5$ (nuclear matter)
and $Z/A = 0$ (neutron matter) are unaffected by the sixth invariant.
The binding energy curves for intermediate values of $Z/A$ will lie between these
two extreme curves, which does not provide a large freedom. 
A more quantitative measure of the equation of state is the asymmetry-energy 
coefficient $a_4$, multiplying the $4 (Z-A)^2/A$ term 
in the semi-empirical mass-formula.
As has been argued in \cite{terHaar_asym} one should use the 
binding energies at the various saturation points in the determination of
this coefficient and we determine $a_4$ by
\begin{equation}
a_4 = \frac{1}{4} \frac{\partial^2}{\partial \xi^2} E_b(\xi) 
\mid_{\xi = Z/A, \rho = \rho_{eq}(Z/A)}.
\end{equation}
Using this prescription we find for the Groningen potential $a_4 = 25$ MeV. 
Recalculating it for the Bonn C potential \cite{Machleidt} we find a 
sightly larger value of 28 MeV.
In the literature one finds a rather wide range of phenomenological values.
Older ones tend to give lower values, e.g. in \cite{Green} a value of 
$a_4 = 23.7$ MeV is given, a more recent liquid drop model calculation cites a
result in the range 27-30 MeV while a very recent work comes up with 
$a_4 = 32.65$ MeV \cite{Myers}.
Our result also shows that extracting the asymmetry coefficient 
via fitting self-energies only for symmetric nuclear and neutron matter,
as is done e.g. in Ref. \cite{Savushkin}, gives much too large results.
These authors find values of 35 MeV and higher.
We also note that although for the binding energy we find a nice
quadratic dependence on the asymmetry parameter, this is not the case
for the other quantities like effective masses and self-energies 
(taken at the respective Fermi-momenta). 
There we find a dependence which is more linear in character.

\begin{figure}
\centerline{\epsfig{file=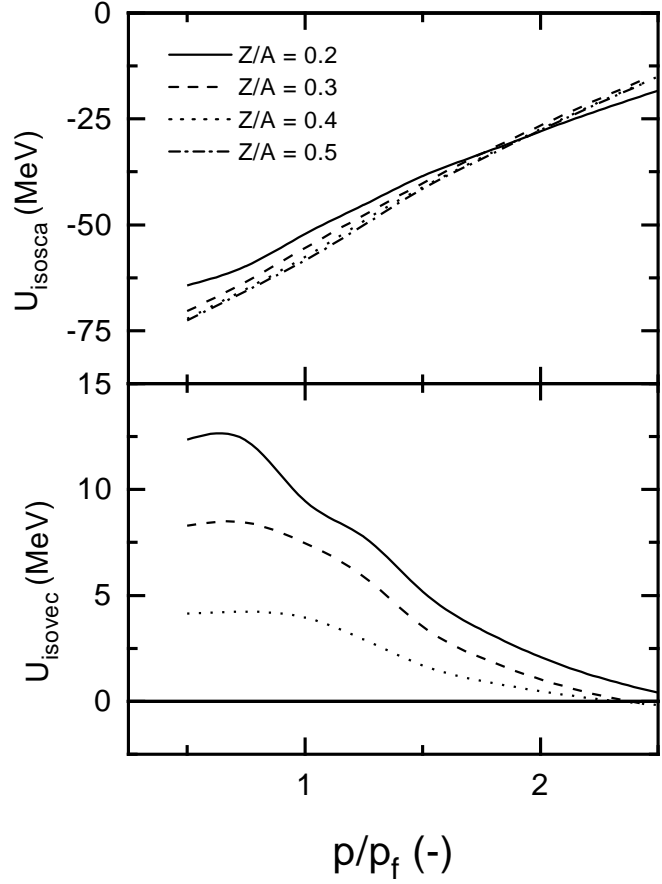,height=20cm}}
\caption{
The iso-scalar and iso-vector components of the mean-field 
for various asymmetry parameters.
The mean-fields are evaluated at the saturation density for the
respective asymmetry parameter.
The momentum is given relative to the average Fermi-momentum as 
defined in the text.
}
\label{Figure_MF}
\end{figure}

In studying the momentum dependence of the self-energies it is convenient to
define isoscalar and isovector quantities. 
For a given observable $O$ we set
\begin{equation}
O_s = \frac{1}{2}(O_n + O_p), 
O_v = \frac{1}{2}(O_n - O_p), 
\end{equation}
In Fig. \ref{Figure_MF} we show the isoscalar
and isovector mean-field as a function of the relative momentum $p/p_f$,
calculated at the respective saturation densities for each asymmetry parameter.
As in Refs. \cite{terHaar,FdJ_spec} we define the mean-field as the shift in the 
pole of the nucleon propagator due to the medium correlations.
We see that up to an asymmetry parameter of 0.3 the isovector mean-field is 
rather independent of $Z/A$. 
There are some minor differences, but the slope is essentially the same.
Only at $Z/A = 0.2$ we find a deviation, this might be due to the fact
that we calculate the mean-field at the saturation density, which is different
in this case. 
The isovector mean-field shows more structure. 
We observe the intuitively expected increasing difference with
increasing asymmetry (corresponding to a lower value of $Z/A$). 
However, there is no quadratic dependence on the density
A significant dependence on the momentum is found for high momenta,
while inside the Fermi-seaa smooth behaviour is obtained.

We show a similar analysis for the isovector 
components of the scalar and vector self-energy. 
The isovector scalar and vector self-energies can be interpreted as a
`effective' $\delta$-meson and $\rho$-meson strengths.
Again we see a notable momentum dependence, similar to the one
observed in the isovector mean-field.

It is also interesting to analyze our results in terms of mean-field
effective coupling constants. 
This is particulary useful in calculations of finite nuclei \cite{DDRH}.
As in the analysis above, we have four channels, both a 
scalar and vector in either the isoscalar and isovector channel.
For this analysis we ignore the momentum dependence and
calculate the self-energy of both the neutron and proton  
at the average Fermi-momentum as defined in Eq. \ref{fermi_mom}. 
The effective coupling constants are then defined by:
\begin{eqnarray}
\left(\frac{g^*_{\sigma}}{m_{\sigma}}\right)^2 
&=& -\frac{1}{2} 
\frac{\Sigma^{s}_{n}(\bar{p}_f) + \Sigma^{s}_{p}(\bar{p}_f)}{\rho_n^s + \rho_p^s }
\nonumber\\
\left(\frac{g^*_{\omega}}{m_{\omega}}\right)^2 
&=& -\frac{1}{2} 
\frac{\Sigma^{0}_{n}(\bar{p}_f) + \Sigma^{0}_{p}(\bar{p}_f)}{\rho_n^v + \rho_p^v}
\nonumber\\
\left(\frac{g^*_{\delta}}{m_{\delta}}\right)^2 
&=& -\frac{1}{2} 
\frac{\Sigma^{s}_{n}(\bar{p}_f) - \Sigma^{s}_{p}(\bar{p}_f)}{\rho_n^s  - \rho_p^s }
\nonumber\\
\left(\frac{g^*_{\rho}}{m_{\rho}}\right)^2 
&=& -\frac{1}{2} 
\frac{\Sigma^{0}_{n}(\bar{p}_f) - \Sigma^{0}_{p}(\bar{p}_f)}{\rho_n^v - \rho_p^v},
\end{eqnarray}
where $\rho^s_i$ and $\rho^v_i$ are the usual scalar and 
vector densities from relativistic mean-field theory \cite{SW}.
The results are presented in Fig. \ref{Figure_eff_g}. 
As one intuitively expects, the scalar and vector isoscalar channels carry 
the largest strength. 
The dependence on the asymmetry parameter is small but we do observe a 
strong dependence on the density. 
For the densities presented in the figure, which range from half to
double normal nuclear matter density, the dependence is predominantly 
linear in the Fermi momentum except for the isovector scalar channel.
There, a significant dependence on higher orders in $p_f$ and the 
asymmetry is found. 
This is in agreement with the finding of Boersma and Malfliet for 
their density dependent parametrization of the G-matrix. 
They strongly favour a density dependence of the coupling constants
linear in $p_f$ as well \cite{Boersma}, over
other possibilities like the exponential dependence of Marcos {\it et al.} \cite{Marcos}. 
The strength in the isovector channels is smaller although equal in both 
the scalar and vector channel.
The significant scalar strength in the isovector channel suggests that for 
fits using only a limited number of meson, like the one of Ref. \cite{Marcos}
one certainly needs to include a $\delta$-meson as well.
For densities corresponding to $p_f > 0.24$ GeV/c we again find negligible 
dependence on the asymmetry parameter and a linear dependence on the Fermi
momentum.
We find a positive sign for ${g_\delta^*}^2$, which is not the case
when one uses the momentum dependence of the single-particle energy 
to extract the scalar and vector self-energy components \cite{Ulrych}.
Note that since our decomposition of the self-energy is not
unique the same holds for this analysis in terms of effective coupling
constants. 
However, we think our approximate method still will be able to 
indicate trends.
Still, both the neutron and proton self-energy will have an `error' $\alpha_{n,p}$. 
If these are very different due to e.g. a strong dependence on the momentum, 
one can obtain spurious results in the isovector channels 
as the negative sign of ${g_\delta^*}^2$ indicates when using the 
momentum dependence method.

Below $p_f \sim 0.24$ GeV/c the density dependence changes and
an enhancement of the isovector couplings is found at large asymmetry.
Most probably this is related to the appearance of bound states related to 
pairing,
which typically appear at low densities in a Brueckner approach.
In that sense the Brueckner scheme is an intermediate-density approximation,
losing its physical significance at low densities.
At low densities the Brueckner independent-pair assumption ceases to be valid
because of the onset of the pairing instability and the approach of the deuteron 
bound-state pole in the $np$-channel. 
For $Z/A = 0.2$ the proton density is already very low, and one starts to see
the onset of pairing correlations. 
A similar behaviour can be observed in the density dependent parametrization
of the former results of Ref. \cite{terHaar_asym} by Marcos {\it et al.} \cite{Marcos}.

The widely used relativistic mean-field theories (RMT) \cite{SW,Ring}
are based on Lagrangians including $\sigma,\omega$ and $\rho$ mesons.
Density independent meson-nucleon coupling constants obtained
phemomenologically from fits to nuclear properties are used. 
In Tab. \ref{Table_cpl} our DB coupling constants, taken at the saturation density
$\rho_0$=0.16~fm$^{-3}$, are compared to standard density independent
values from the $\sigma$-$\omega$ model (without scalar
self-interactions) \cite{SW}. For comparison also the density dependent
DB isoscalar scalar and vector coupling constants of ref.\cite{Haddad}
are shown which in \cite{DDRH} were found to describe nuclear properties
very well. In the isoscalar channels the two DB sets are in rather good
agreement and both calculations reproduce almost perfectly the purely
phenomenological RMT coupling constants. 
It is also worth noting that in the isoscalar channel the effective 
coupling constants at saturation density are remarkably close to 
the `bare' values as they appear in the interaction. 
This means, that around these densities the higher order effects generated 
by the Brueckner ladder are small.
We therefore should find only small dependencies on e.g. the asymmetry, which
we do observe.
In the isovector channel the situation is totally different, there the effective
coupling constants are much larger than the bare values. 
This implies that the (density and asymmetry dependent) higher order effects 
due to the Brueckner ladders are significant. 
Indeed, we see a pronounced dependence on asymmetry and density of the 
effective coupling constants of the effective isovector mesons.

Most important and conclusive for field theoretical models are the
results for the isovector channels. The DB and RMT $\rho$ coupling
appear to be surprisingly close but is has to be noted that the
isovector RMT coupling constants are not well determined. In fact, there
exist other RMT parameter sets \cite{SW} obtained from fits to finite
nuclei with slightly different isoscalar couplings but about twice as
large values for g$^2_\rho$. From Tab. \ref{Table_cpl} it is seen that the
isovector-scalar and the isovector-vector DB couplings are of
comparable strength. This result gives clear evidence that the
$\delta$-meson should be included in effective field theories. 
The scalar coupling of the $\delta$ meson has important consequences for the
dynamics in N $\neq$ Z systems. Because protons and neutrons will carry
different effective masses they become dynamically distinguishable while
the $\rho$ meson leads primarily to energy shifts. Most important for
applications to finite nuclei is that isovector effects in the
spin-orbit potential will be enhanced as required by single particle
spectra in charge asymmetric nuclei. Inclusion of the $\delta$ meson
might help to resolve the uncertainties in the RMT isovector sector.

\section{Conclusions}

We presented a calculation of asymmetric nuclear matter in a 
relativistic Brueckner framework. 
To extract the components of the self-energies we used the 
method of projecting on Lorentz-invariant amplitudes.
We found that for asymmetric nuclear matter a sixth invariant is needed
on top of the conventional set of five invariants as for example used in 
Refs. \cite{terHaar,FdJ_spec}.
This sixth invariant has to be chosen in such a way that for 
an asymmetry parameter close to $0.5$ its coefficient 
has to vanish so that one recovers the usual description of
symmetric nuclear matter in terms of five Lorentz-invariant
amplitudes. 
We presented a choice of this sixth invariant that satisfies
this criterium.
With this model we performed calculations for several observables.
The binding energy of the saturation point showed the expected
quadratic dependence on the asymmetry parameter.
For the asymmetry energy we found a value of around 25 MeV.
We analysed our results in terms of mean-field scalar-vector isoscalar-isovector
quantities. 
Apart from the expected scalar-vector strength in the isoscalar channel we
also found significant scalar strength in the isovector channel. 
This can be interpreted as an effective $\delta$-meson, which couples as 
strong as the effective isovector-vector $\rho$-meson.
Finally, we found a density dependence linear in the Fermi momentum of the 
effective coupling constants, which was also observed in other calculations
\cite{Boersma}.

\newpage

\newpage

\begin{figure}
\centerline{\epsfig{file=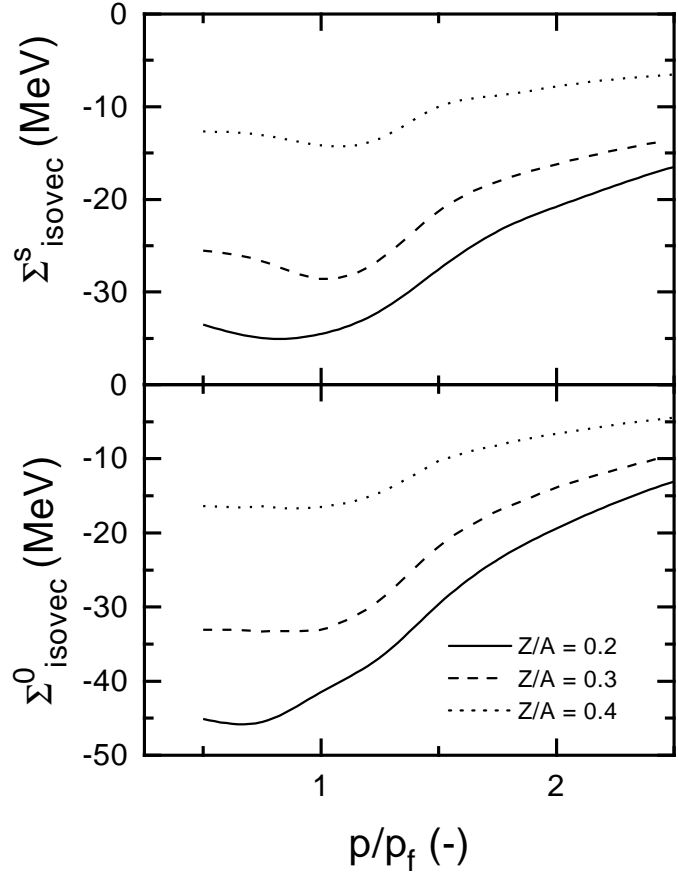,height=20cm}}
\caption{
The isovector components of the scalar and vector components
of the self-energy for various asymmetry parameters.
These are evaluated at the saturation density for the
respective asymmetry parameter.
The momentum is given relative to the average Fermi-momentum as 
defined in the text.
}
\label{Figure_sigs}
\end{figure}

\newpage

\begin{figure}[h]
\centerline{\epsfig{file=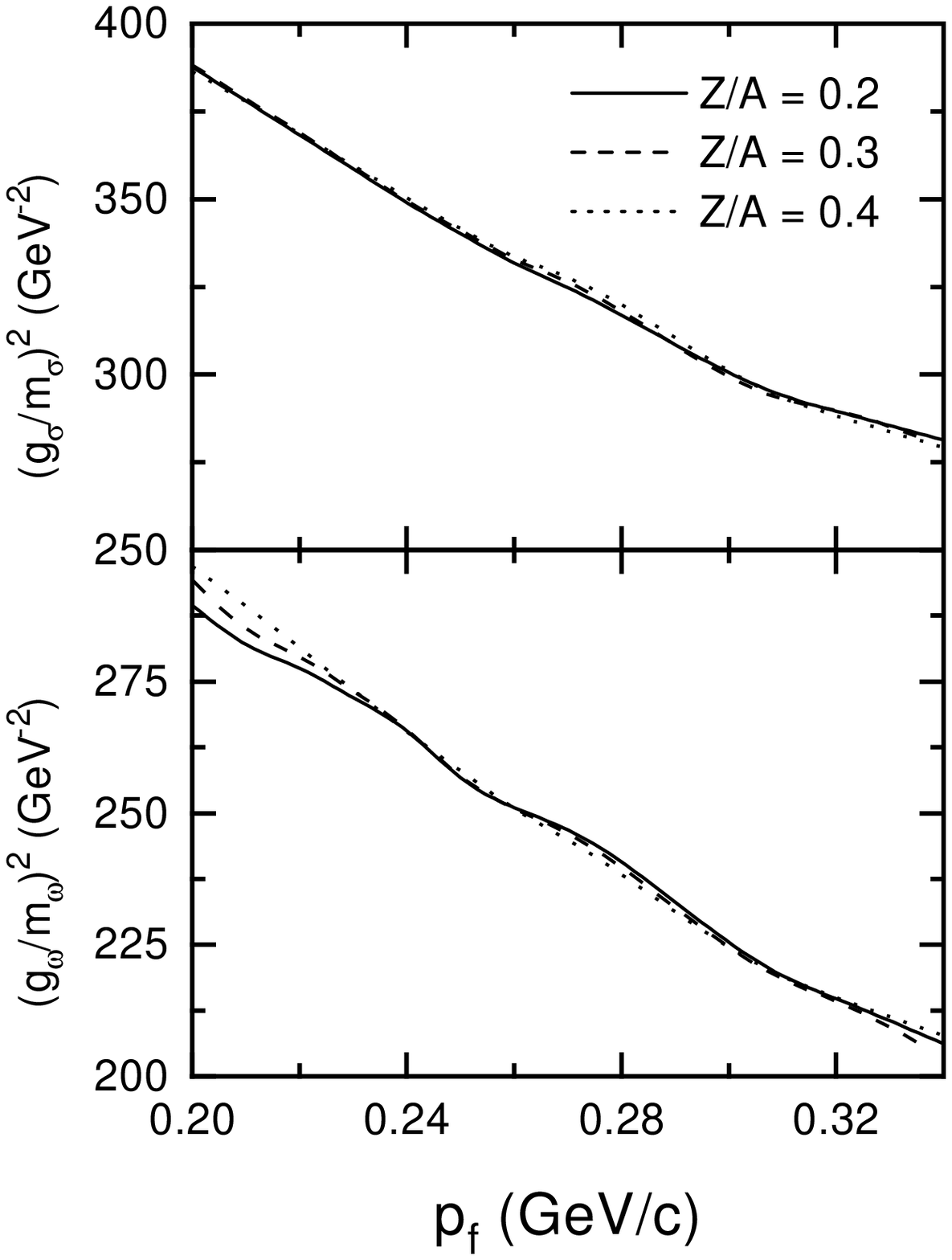,height=20cm}}
\centerline{\epsfig{file=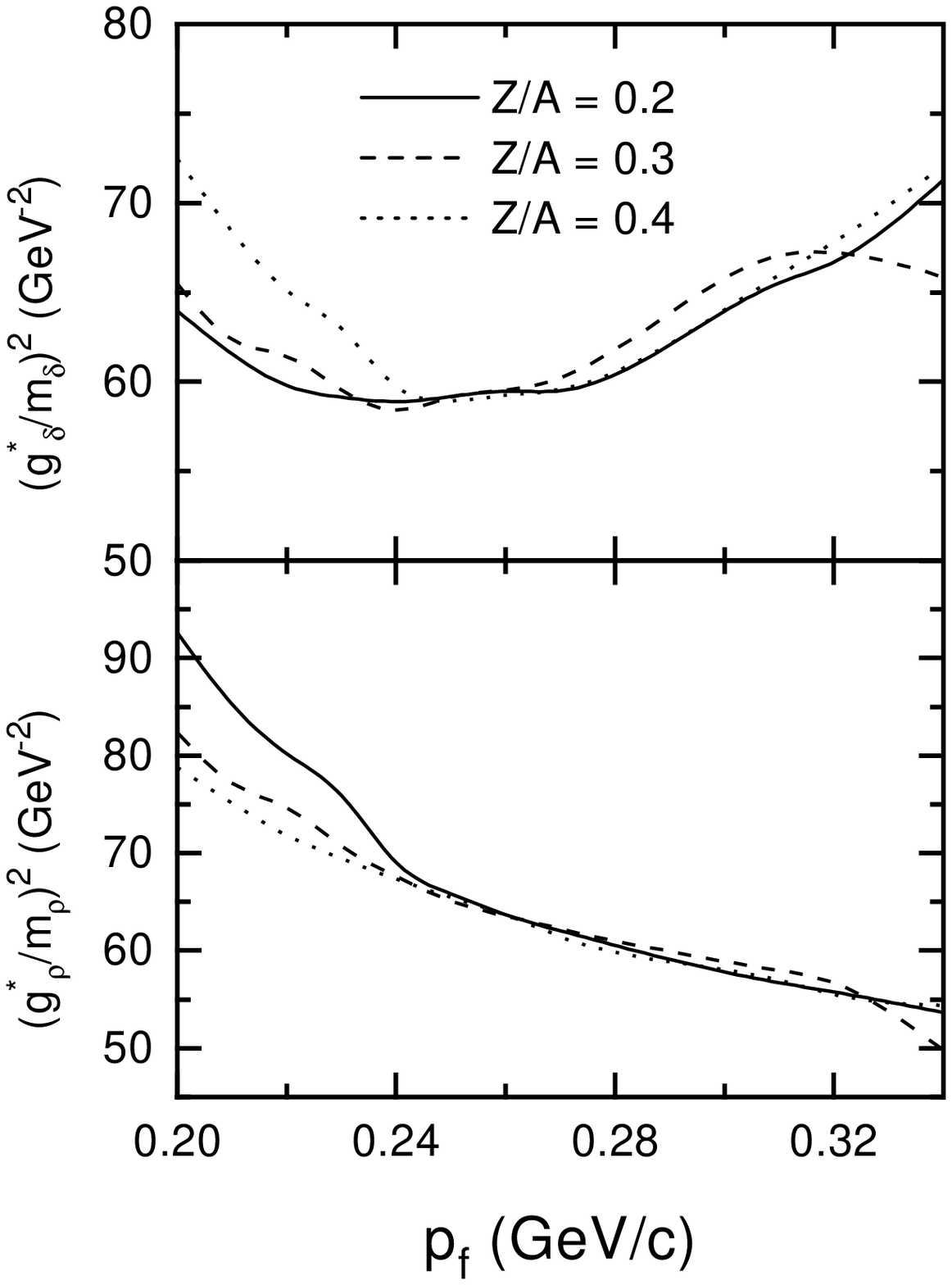,height=20cm}}
\caption{
The effective coupling constants in the isoscalar scalar ($g_{\sigma}$),
and vector ($g_{\omega}$) and the isovector scalar ($g_{\delta}$)
and vector ($g_{\rho}$) as defined in the text for various asymmetry 
parameters and densities.
}
\label{Figure_eff_g}
\end{figure}

\begin{table}
\begin{center}
\begin{tabular}{ccccc}
Meson & g$^2_{DB}/4\pi$ & g$^2_{HW}/4\pi$ & g$^2_{RMT}/4\pi$ & Mass~[MeV]\\[0.5ex] \hline
$\sigma$  &  7.82   &  7.51     &  7.29     & 550  \\
$\omega$  & 12.10   & 11.52     & 10.84     & 783  \\
$\rho$    &  2.97   &   -       &  2.93     & 770  \\
$\delta$  &  4.61   &   -       &   -       & 983  \\
\end{tabular}
\end{center}
\caption{DB coupling constants taken at saturation density
$\rho$=0.16~fm$^{-3}$ of the present work (first column) and from 
ref.\protect\cite{Haddad} (second column, obtained from the
Bonn A NN-potential) are compared to
RMT values \protect\cite{SW} (third column). The meson masses
used to derive the coupling constants from the results of Fig.2 are
displayed in the last column.
}
\label{Table_cpl}
\end{table}

\end{document}